\documentclass[aps,prl,twocolumn,superscriptaddress,showpacs]{revtex4}
\usepackage[normalem]{ulem}
\usepackage{float}
\usepackage[usenames]{color}
\usepackage{graphicx}
\usepackage{ulem}
\usepackage{amsfonts}
\usepackage{amsmath}
\begin{document}

\title{Nonthermal Fluctuations of the Mitotic Spindle}

\author{K. Smith}
\affiliation{Department of Physics, University of Massachusetts Amherst, Amherst, Massachusetts 01003, USA}
\author{B. Griffin}
\affiliation{Department of Physics, University of Massachusetts Amherst, Amherst, Massachusetts 01003, USA}
\author{H. Byrd}
\affiliation{Department of Physics, University of Massachusetts Amherst, Amherst, Massachusetts 01003, USA}
\author{F. C. MacKintosh}
\affiliation{Department of Physics and Astronomy, VU University, Amsterdam, The Netherlands}
\author{M. L. Kilfoil}
\affiliation{Department of Physics, University of Massachusetts Amherst, Amherst, Massachusetts 01003, USA}

\pacs{87.16.Ka, 87.14.gk, 87.16.dm}
\date{\today}

\begin{abstract}
We present direct measurements of fluctuations in the nucleus of yeast cells.
While prior work has shown these fluctuations to be active and non-thermal in character, 
their origin and time dependence are not understood. 
We show that the nuclear fluctuations we observe are quantitatively consistent with 
uncorrelated, active force fluctuations driving a nuclear medium that is dominated by 
an uncondensed DNA solution, for which we perform rheological measurements on 
an {\it in vitro} model system under similar conditions to what is expected in the nucleus.
\end{abstract}

\maketitle
\noindent
Living cells necessarily operate far from equilibrium.
The cytoplasm is a highly dynamic medium, with incessant energy-consuming active processes and motion. 
Much of this activity arises from force generation by {\it molecular motors}. 
While much is known about these motor proteins from single-molecule studies {\it in vitro}, as well as from fluorescence imaging that can track individual motors {\it in vivo} on the nanometer scale, we still know rather little about the collective effects of motor activity at the mesoscopic scale, especially in the nucleus. 
Recent studies have identified a prevalent form of intracellular motion that has sometimes been called {\it active diffusion}, which is non-directed yet non-thermal motion that can often appear similar to thermal Brownian motion~\cite{BranFCMWeitzJCB08}. This often has an amplitude that can be much greater than the thermal motion and is expected to have a different time dependence.
Given the importance of such motion for transport within the cell, many of the recent studies have focused on ways to identify the non-equilibrium origins of motion in cells. 
This has proven to be experimentally very challenging, with most results 
{\it in vivo} being indirect~\cite{Lau-PRL03,BranFCMWeitzJCB08,Daisuke-Christoph-beadcells-2009,Fakhri2014}, often employing qualitative measures such as drug interventions or ATP depletion to suppress motor activity. By contrast, direct active microrheology approaches have been limited to {\it in vitro} reconstituted systems~\cite{Daisuke-Science-2007}, up until similar methods were recently employed
to directly measure the spectrum of non-equilibrium fluctuations in the cytoplasm of living cells~\cite{Guo-2014}.

Non-equilibrium activity on smaller scales, in both bacteria and cell nuclei, have been identified using 
a combination of imaging and drug intervention.
Several measurements of the motions of chromosome loci, employing an inserted array of fluorescently-labeled-protein binding sites on the chromosome, have been performed to extract the mean squared displacement (MSD) of position fluctuations within living yeast cell nuclei~\cite{Gasser-PNAS04,LarsonBloom-2010,Weber-PNAS-2012}. Sub-diffusive motion of individual chromosome loci using two-dimensional imaging, as well as of the relative motion of two loci on the same chromosome using three-dimensional imaging, have been reported~\cite{Gasser-PNAS04,LarsonBloom-2010}. Recently, Weber {\it et al.}~\cite{Weber-PNAS-2012} used 2D microscopy and drug interventions to show that this sub-diffusive motion was non-thermal in origin. These authors, however, were unable to identify the origin of these fluctuations, and they suggested that the athermal activity exhibited a time dependence similar to thermal forces. 

Here, we combine a recently-developed method of high resolution three-dimensional microscopy and tracking with two-point microrheology to study fluctuations in the eukaryotic nucleus and mitotic spindle.
We show that the observed fluctuations are quantitatively consistent with uncorrelated force fluctuations in dense solutions/gels of DNA, for which we perform rheological measurements on an {\it in vitro} model system under similar conditions to what is expected in the nucleus.
In particular, we show that although the time dependence of position fluctuations is consistent with prior observations, the implied 
force fluctuations are far from what is expected for thermal motion. This is in contrast to the interpretation given by Weber {\it et al.}~\cite{Weber-PNAS-2012}, but is consistent with a simple model of uncorrelated active fluctuations in dense DNA solutions governed by the Zimm model for flexible polymer dynamics \cite{MacKintosh-PNAS-2012}. 

\begin{figure}
\begin{center}
\includegraphics[width=0.85 \columnwidth]{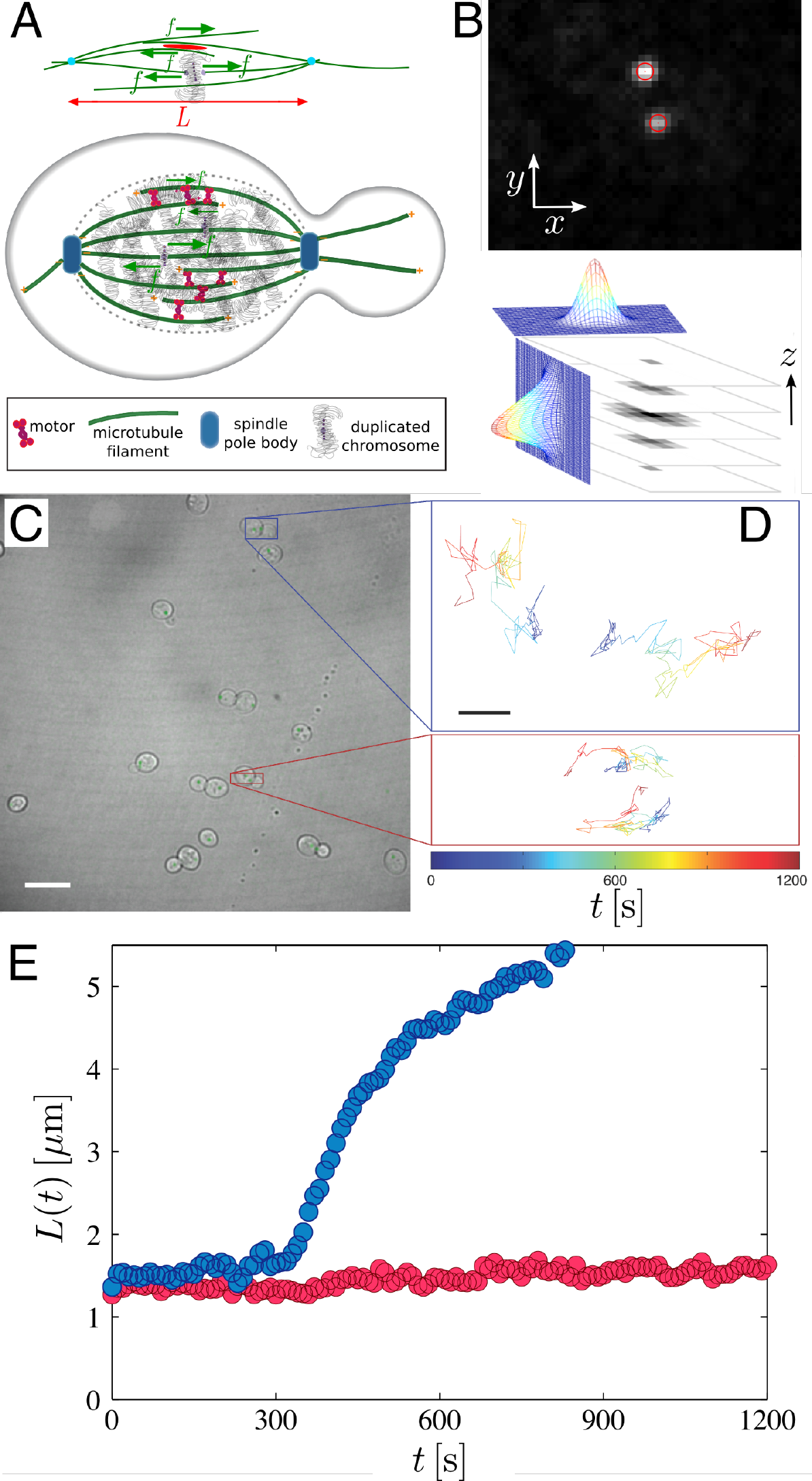}\vspace{-0.15in}
\caption{\label{fig:invivo_system_methmat} a) Schematic diagram of active force fluctuations and the eukaryotic nuclear environment. In the spindle mid-zone, coherent motion of molecular motors (red) or (de)polymerization of microtubule filaments (green), drive the network filaments, generating forces (green arrows) in a viscoelastic medium. b) Illustration of 3D subpixel-resolution position-finding by fitting a 3D Gaussian function to the full intensity distribution (scale in image data: $0.185\,\mu\mathrm{m}$/pixel).  c) Composite image of the raw fluorescence intensities from one confocal image stack with bright-field image for a typical  field of view for an unsynchronized population of wild-type cells (scale bar: $10\,\mu\mathrm{m}$). d) Pole trajectories obtained by analysis of the 3D image time series  for the two cells highlighted in c), projected into $x-y$ plane (scale bar: $1\,\mu\mathrm{m}$). e) Time-evolution of the 3D pole-pole separation $L(t)$ for each cell in d). 
Cell  in red remains in metaphase and exhibits fluctuations. Cell in blue undergoes anaphase chromosome separation.}
\end{center}\vspace{-0.3in}
\end{figure}

Before cell division, in what are known as {\it metaphase} and the subsequent {\it anaphase}, when chromosomes are separated, the eukaryotic nucleus is spanned by the mitotic spindle, consisting of an intricate network of protein filaments, microtubules (MTs), self-organized to originate from nucleating centers and terminate either at connections to the chromosomes or in an anti-parallel alignment in the spindle mid-zone. This filament network is both stabilized and driven by specialized motor proteins that slide antiparallel microtubules relative to one another, converting the chemical energy of adenosine triphosphate (ATP) to mechanical work and motion. At the same time, the MTs are themselves dynamic and their incessant polymerization and depolymerization can also introduce non-equilibrium forces into the spindle~\cite{DogteromMTforces,MitchisonRef}, although these forces are not directly ATP-dependent. These forces are transmitted to the surrounding micro-mechanical environment of the eukaryotic nucleus, likely dominated by the DNA that is prevalent throughout the nucleus~\cite{StraightMurray-Science97}. The spindle and forces transmitted to the surroundings are sketched in Fig.~\ref{fig:invivo_system_methmat} (a). 

The stochasticity of both motor and polymerization forces in the spindle can drive non-thermal fluctuations in the nucleus. It is not known, however, which of these forces, if either, dominates spindle fluctuations. 
To measure these fluctuations, we developed methods to track the spindle dynamics of single cells in three dimensions. 
We used the model eukaryote {\it Saccharomyces cerevisiae} to investigate active motion in the nucleus in detail. 
To quantify spindle length, $L$, we labelled the spindle-pole body protein Spc42 with green fluorescent protein (GFP) and measured the separation between the poles. 
We used confocal microscopy to carry out three-dimensional time-lapse imaging at sub-pixel resolution on living cells. 
This organism has a genome length of 12 Mb comprising 16 chromosomes in the haploid cell, and undergoes a mitosis every 90 minutes under optimum growth conditions. Rather than being highly condensed, the DNA is dispersed throughout the nucleus even during metaphase~\cite{StraightMurray-Science97,Cross-irreversibleCellCycleStart}.  

All yeast strains used are derived from strain BY4741 \cite{Boekeyeaststrain}. 
All cells were cultured and imaged in synthetic complete (SC) minimal medium~\cite{CSHL_manual} to minimize adverse effects from photon absorption on the natural behavior. 
The SPB reporter Spc42-GFP is described in \cite{StraightMurray-Science97}. For each type of analysis, data was collected from cell populations representing four independently derived yeast strains. 

For the motor mutants, the Cin8 and Kip1 deletion strains were tested in several different ways, using genetic and genomic (PCR) methods, as described in the accompanying ESI.
Taken together, the PCR results verify that the deletions are present and that there is no wild type copy of either Kip1 or Cin8 present; and the genetic crosses verify that these strains are synthetically lethal. 

For imaging, cells were grown overnight at $30^{\circ}$C, diluted in SC and incubated for 2 hours to allow re-initiation of log-phase growth, gently centrifuged and resuspended at the desired density  of $7\times 10^8\,\mathrm{cells/mL}$. $8\,\mu\mathrm{L}$ of this culture were transferred onto a freshly-prepared SC-agar pad and covered with a glass coverslip. 
For cells treated with metabolic inhibitors, standard sodium azide and 2-deoxyglucose treatment was used~\cite{KellerBusmotors}.
Cells were imaged at 30\,$^{\circ}$C in stacks of 21 focal planes spaced $0.3\,\mu\rm{m}$ apart at 10 s per stack, using a custom multi-beam scanner ($512\times512$ pinholes) confocal microscope system (VisiTech) mounted on a Leica DMIRB equipped with a Nanodrive piezo stage (ASI) and 488 nm solid state laser controlled by an AOTF. Images were acquired with a EM-CCD camera (Hamamatsu) using a 63X 1.4 NA objective.  
From the images, we computed the positions of the spindle pole bodies at three-dimensional sub-pixel resolution~\cite{tracking09} by fitting the intensity distribution to a three-dimensional Gaussian function (Fig.~\ref{fig:invivo_system_methmat} (b)), obtaining a spatial resolution of $<10\,\mathrm{nm}$ in 3D for each pole. All analysis was carried out using custom scripts in MATLAB (MathWorks). 

Cells at random stages of the cell cycle in the asynchronous population were represented in each field of view.  Cells in both metaphase and anaphase have two separate spindle poles that can be easily distinguished in the analysis from cells in other stages of the cell cycle. From the trajectories of both poles, the time-dependent length of the mitotic spindle, $L(t)$, was determined for this subset of cells. While visual examination of the trajectories of each pole for metaphase cells shows apparently random motion (Fig.~\ref{fig:invivo_system_methmat} (d)), the 3D separation between the poles shows coherent dynamics. 
This approach has the powerful advantages of allowing unambiguous discrimination between true spindle length fluctuations, which are the quantities of interest here, and out of plane spindle excursions; by virtue of the 3D imaging and high resolution; and segregation of bipolar spindle cells, and specifically those in metaphase, from a population of normally-growing cells, without resorting to addition of a drug to stall growth in one phase of the cell cycle.  
The rotationally invariant measure $L(t)$ is immune to any advection of the whole cell or of the microscope stage, or of the entire nucleus, and to instability of the focal plane and other such physical artefacts that are possible contaminants of fine measurements of individual locus motions.

We observed typically 12-15 cells per field of view (see Fig.~\ref{fig:invivo_system_methmat} (c)), from which 2-3 cells were in metaphase over the 20 minute observation window. Data were collected over many fields of view under identical preparations and growth conditions. 
We show $L(t)$ for a representative cell in each of these two phases in Fig.~\ref{fig:invivo_system_methmat} (e). Cells in metaphase (red) have an approximately steady-state spindle length $L(t)$, as observed previously for budding yeast cells~\cite{StraightMurray-Science97}, while cells in anaphase (blue) exhibit a clear increase in spindle length. 
The data show fluctuations in $L(t)$ during metaphase that are easily distinguishable above the 3D resolution in the position localization. 

\begin{figure}
\begin{center}
\includegraphics[width=0.95 \columnwidth]{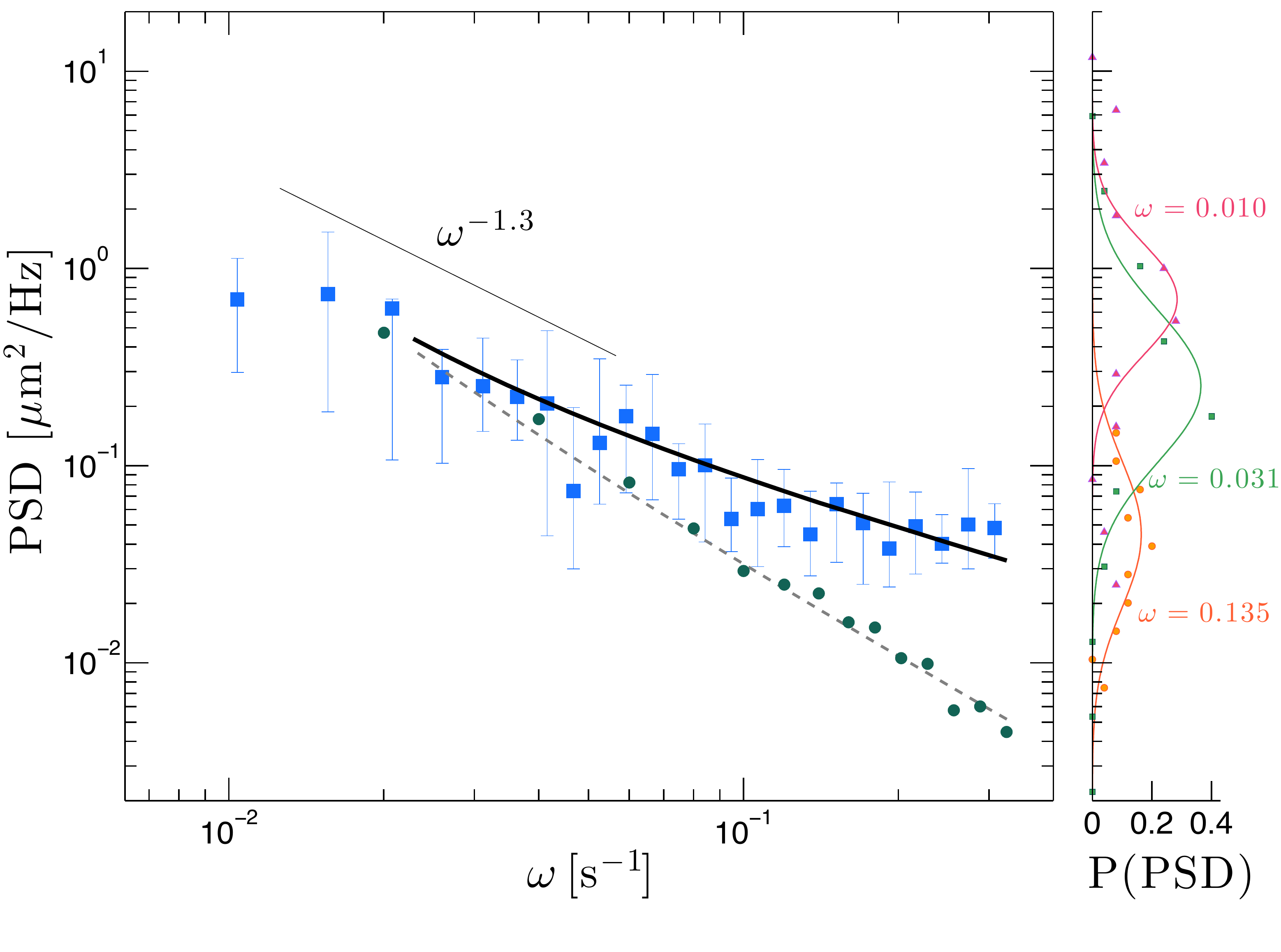}\vspace{-0.15in} 
\caption{\label{fig:psd} Experimentally-determined displacement power spectral distribution (PSD) from living cell nucleus determined for 25 separate metaphase cells. The peak of the lognormal distribution of the data at each frequency is plotted (square symbols), with the first and third quartiles indicated by  error bars. The observed spectrum follows an approximate power law $PSD \sim \omega^{-1.3}$ over the longest timescales (indicated by long-dashed line). Theory curve for the prediction for active-white noise (solid line) is plotted superimposed on the data, with arbitrary multiplicative factor. The circles and dashed line represent the thermal PSD obtained directly from passive beads and from the approximated $|G|$ using fluctuation-dissipation theorem, respectively, from the {\it in vitro} experiment in the absence of active fluctuations; plotted with arbitrary multiplicative factor. 
At right: the power across the cell population was distributed log-normally at each frequency. Fits are shown for three frequencies spanning the spectrum.}
\end{center}\vspace{-0.3in}
\end{figure}
To explore these fluctuations in the metaphase cells' spindles, we collected the $L(t)$ data for a population of cells, and from the fluctuations of $L(t)$ we calculated the displacement power spectral density (PSD) by Fast Fourier Transform~\cite{ChristophPSD98}. 
As the power spectral density analysis assumes uniformly spaced time series, we excluded cells that lacked the position for a single time point of either pole. In all, 25 cells were included in the analysis. 
The PSD data for cells are presented in Fig.~\ref{fig:psd}. The ensemble-averaged power spectrum shows an approximate power-law behavior in frequency $\omega$, with the PSD$\sim \omega^{-1.3}$\ at low frequency, 
although the data cannot be captured by a single power-law throughout the experimental range.
This behavior can be understood by a variation of models for active diffusion in eukaryotic cells~\cite{Lau-PRL03,ActiveGelThyPRL,BranFCMWeitzJCB08}. Specifically, our observations are consistent with uncorrelated active force fluctuations driving motion in a viscoelastic medium~\cite{MacKintosh-PNAS-2012}. 

Such uncorrelated force fluctuations are expected for the long time-scales of our experiments. Correlations can arise, for instance, from the {\it processivity} or coherent motion of molecular motors or microtubule polymerization and depolymerization. 
The frequency range we observe corresponds to time-scales ($2\pi/\omega\gtrsim15\,$s) that are longer than the 
correlation time of either of these processes ($\lesssim5\,$s)~\cite{modelGardnerOdde,VeigelSchmidt-review_motors-2011,Khmelinskii-DevCell-2009}. 
Thus, the active forces should be  uncorrelated~\cite{ActiveGelThyPRL,Levine-FCM-JPC09} and the force spectrum should be white. 
This is in contrast to thermal driving forces that must be correlated in viscoelastic systems, as a consequence of the fluctuation-dissipation theorem~\cite{Landau}.
In a medium with power-law rheology given by $G(\omega)\sim\omega^\alpha$, the thermal MSD$\,\sim t^\alpha$, corresponding to a PSD$\,\sim \omega^{-(1+\alpha)}$. By contrast, for white noise force spectrum, the MSD$\,\sim t^{(2\alpha-1)}$ and the PSD$\,\sim \omega^{-2\alpha}$.
For more a general frequency-dependent modulus $G(\omega)$, when driven by a force $f$, the frequency-dependent displacement $x$ should vary inversely with the stiffness of the matrix: $x\sim f/G$. Thus, for white noise driving, the power spectrum of position fluctuations $\langle x^2(\omega)\rangle\sim1/|G(\omega)|^2$.

There have been very few reports of directly measured nuclear rheology in living cells, and none for the cells we study here.
Measurements of the fluctuations of the nuclear envelope labelled of a nuclear pore complex protein, do not account for the relaxation spectrum (data not shown).
To investigate the viscoelastic environment in the nucleus, we created an {\it in vitro} model system of bare chromosomes at a DNA density 
comparable to the nuclear environment. 
We then performed passive microrheology measurements~\cite{masonweitz,GittesPRL97,Schnurr1997,microrheoDasguptaWeitz,microrheo09} on this model system. 
The model system consisted of $1\,\mathrm{mg/ml}$ $\lambda$-DNA (48,502 b.p.) in physiological salt conditions.
Under these conditions, the DNA is well above the semi-dilute limit~\cite{Pernodet_lambdaDNA_xover97} and can be considered an entangled polymer network. 
$\lambda$-DNA (New England Biolabs) was concentrated by ethanol precipitation then resuspended in buffer (10 mM Tris-HCl, pH 7.9, 50 mM NaCl, 50 mM KCl, 5 mM $\mathrm{MgCl_2}$, 0.1 mM EDTA, and 15 mg/L bovine serum albumin (BSA)) at the desired concentration, with slow mixing. 
Concentration was measured using a ND-1000 Nanodrop in serial dilutions with accuracy of $<3\%$ error on the mean in 6 measurements. 
Latex microspheres of $0.5\,\mu\mathrm{m}$ radius (Polysciences) were incorporated in the sample for microrheology. The sample was imaged on a Leica DM-IRB inverted microscope in brightfield mode using a 63X 1.4 NA oil immersion objective lens using a CCD camera (Hamamatsu) at 16 fps for 5 min. Data obtained for multiple fields of view were analyzed separately and combined.

\begin{figure}
\begin{center}
\includegraphics[width=0.95 \columnwidth]{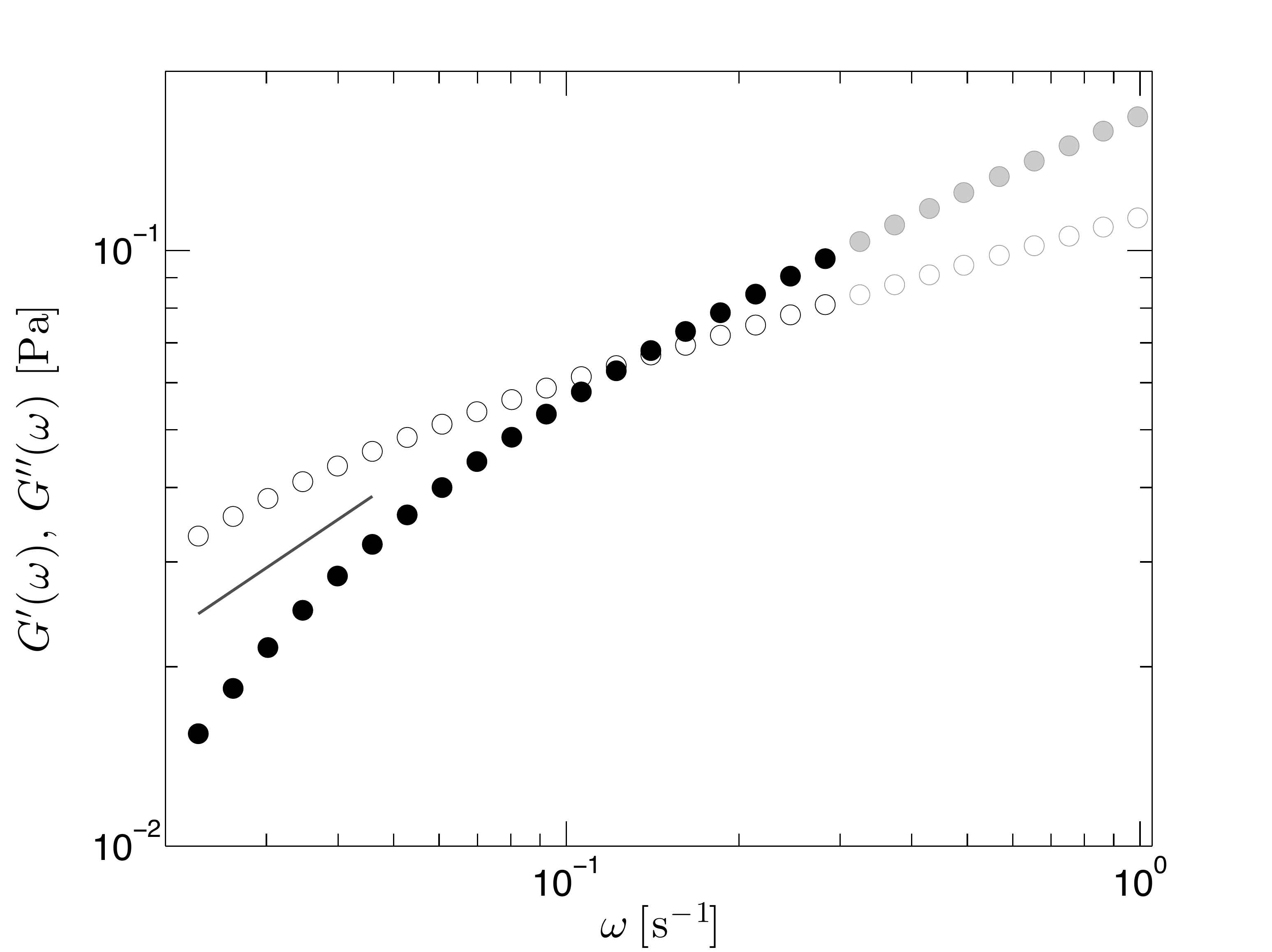}\vspace{-0.15in}
\caption{\label{fig:invitro} {\it In vitro} rheology of a close approximation to the  {\it in vivo} nuclear microenvironment in terms of DNA alone. Viscoelastic moduli $G^{\prime\prime}(\omega)$ (open symbols) and $G^{\prime}(\omega)$ (filled symbols) obtained by 1pt microrheology are shown, highlighted in black  for the relevant frequency range. The complex modulus $|G(\omega)|$ scales as a power law $|G(\omega)|\sim \omega^{0.65}$  over the lowest frequency range (indicated by solid line).}
\end{center}\vspace{-0.3in}
\end{figure}

The frequency-dependent viscoelastic moduli obtained from this approach are plotted in Fig.~\ref{fig:invitro}, with the relevant frequency range for the {\it in vivo} measurements highlighted. 
The measured rheology over the lowest frequency range is approximated by $|G|\sim \omega^{0.65}$, which is close to the Zimm behavior expected for flexible polymer solutions~\cite{Larson_book}. 
Among the very few reports to date of nuclear rheology measurements on living cells, Ref.\ \cite{Discher-nucleusZimm07} measured $G(\omega)$ of nuclei of human stem and differentiated cells. When these authors studied cells deficient in nuclear lamins, which are not present in our yeast cells, they found results consistent with the $|G|\sim \omega^{0.65}$ that we find in our reconstituted system.
As noted above, the full frequency dependence of the PSD for active driving should follow $1/|G(\omega)|^2$.
For the $G(\omega)$ measured in Fig.~\ref{fig:invitro}, this frequency dependence is in excellent agreement with our measured fluctuation spectrum in Fig.~\ref{fig:psd}, as shown by the solid black line.
Here, we do not know the overall strength of active fluctuations, which should depend on such variables as the density and force level of active processes. So, only the amplitude (vertical position in the log-log plot) has been adjusted, and there are no further free parameters.

The implied force fluctuations in this picture are far from what is expected thermally. For a continuum viscoelastic medium characterized by a complex modulus $G(\omega)$, the displacement PSD for a particle of radius $a$ is given by $2k_BTG^{\prime\prime}/\left(6\pi\omega a |G|^2\right)$~\cite{GittesPRL97,Lau-PRL03,Landau}, for which the force spectrum should be proportional to $G^{\prime\prime}(\omega)/\omega$. In general, this represents a colored spectrum. For instance, the approximate $G(\omega)\sim \omega^{0.65}$ indicated in Fig.~\ref{fig:invitro} would correspond to a thermal force spectrum $\sim\omega^{-0.35}$, which is obviously very different from the expected active, white spectrum. 
For a more direct comparison with our measured PSD, we also show on Fig.~\ref{fig:psd} the thermal PSD of probe particles in our {\it in vitro} reconstituted DNA solution. Here, too, the overall amplitude is not known, and a single multiplicative factor has been adjusted for comparison. The data clearly identify quantitative differences between active and thermal driving.
The fact that the latter does not account for the PSD {\it in vivo} suggests that the active driving is dominant and has a different time/frequency dependence than thermal, unless the rheology of the nucleus is very different from our {\it in vitro} model.
But, in any case, for viscoelastic media such as DNA solutions, the active and thermal PSD of fluctuations are very unlikely to be the same~\cite{Weber-PNAS-2012}: 
one implication of the fluctuation dissipation theorem is that the thermal forces in such media must be characterized by a broad spectrum of characteristic timescales, in contrast with such active processes as the motion of molecular motors that exhibit a well-defined processivity time. 

Our results demonstrate that position fluctuations within the nucleus of cells can be fully accounted for by a simple model of a viscoelastic medium, whose response is dominated by a solution of uncondensed DNA, driven by de-correlated active force fluctuations \cite{MacKintosh-PNAS-2012}. One consequence of such active force fluctuations would be enhanced transport relative to thermal fluctuations, at least over a range of time scales of order seconds. 

While the fact that fluctuations in the nucleus on such time scales are non-thermal may not necessarily be surprising in itself, the nature of activity within the nucleus remains almost entirely unknown. In this system, the two most likely sources of activity are molecular motors and (de)polymerization forces due to assembly/disassembly of microtubules. 
In order to test for the former, we individually suppressed each of the two microtubule-associated motors known to be important here: this organism has two types of kinesin-5 tetrameric motors, the proteins Kip1 and Cin8, that can generate outward-directed forces by sliding antiparallel microtubules with respect to each other. When we remove either of these using complete deletion of the gene, we find no statistically significant change in the observed spectrum (see figure 1S in ESI). We conclude that neither of these motors can be {\it solely} responsible for the observed fluctuations. It is not possible, unfortunately, to knock out both motors simultaneously (though such a scenario  can be achieved by temperature-sensitive mutants) {\it and still have an intact (non-collapsed) metaphase spindle}, as these motors are known to play a crucial role in stabilizing the spindle structure~\cite{SaundersHoyt_kinesinproteins92,SaundHoyt95}. 

Recent work has focused on the mechanisms by which Kip1 and Cin8 motors perform their multiple mitotic roles and their regulation~\cite{Avunie-Masala2011,Khmelinskii-DevCell-2009}. 
Phosphoregulation of the MT-bundler Ase1 by the Cdk1 kinase and the Cdc14 phosphatase (which reverses the activity of the kinase) controls the timing of Cin8 physical association with the mid-zone~\cite{Avunie-Masala2011,Khmelinskii-DevCell-2009}.
The Ase1 protein crosslinks mid-zone MTs for spindle elongation, and keeps the two half spindles connected throughout elongation. 
In metaphase, Ase1 is sustained in the phosphorylated state by Cdk1, and in this state inhibits accumulation of Cin8 on iMTs which prevents premature elongation and collapse of the metaphase spindle. 
During the transition to anaphase, Ase1 is dephosphorylated, and both the bundling protein and the kinesin-5 Cin8 associate more strongly with the newly-stable antiparallel iMT scaffold, indicating that these two (and presumably additional) proteins work co-operatively to 
regulate and 
drive spindle elongation.  
Khmelinskii {\it et al.}, have shown that the Ase1 protein exponentially increases the rate at which Cin8 binds to the iMTs at the mid-zone, while 
Fridman {\it et al.} have shown that Kip1 is concentrated at the mid-zone only when spindles elongate in anaphase~\cite{FridmanGheber2013}. 
Motor efficiency for extensile driving thus increases dramatically at the onset of anaphase.
During metaphase the motors play primarily a structural role and appear to be non-processive.
Our finding that the perturbations shown in the supporting material do not alter the PSD during metaphase is consistent with this picture. 

In order to test for active, possibly non-motor processes, it is possible to deplete the cell of ATP
by treating the cells with both sodium azide, a
well-known metabolic inhibitor that switches off ATP production in the mitochondria, and 2-deoxyglucose, which  inhibits glycolysis.
Together, these suppress motor activity more generally~\cite{KellerBusmotors}. When we do this, we also see no statistically significant change in the observed spectrum. However, these metabolic poisons are known to incompletely deplete ATP~\cite{OzalpOlsen_nanoATPlevelsYeast2010} and they are even less effective in depleting GTP~\cite{Ditzelmuller-ATPdepl83}. 
We conclude that the active fluctuations in our system most likely come from microtubule (de)polymerization forces, which do not depend directly on ATP. However, testing this will require more direct methods for micro-mechanical characterization in the nucleus, which remains highly challenging.

\begin{acknowledgments}
We thank C.F. Schmidt for enlightening discussions. The authors thank the Aspen Center for Physics (supported by NSF Grant No.\ 1066293), where this collaboration was initiated. FCM was also supported by FOM/NWO.

\end{acknowledgments}

%
\widetext
\pagebreak
\begin{center}
\textbf{\large Supplemental Materials: Nonthermal Fluctuations of the Mitotic Spindle}
\end{center}
\setcounter{equation}{0}
\setcounter{figure}{0}
\setcounter{table}{0}
\setcounter{page}{1}
\setcounter{section}{0}
\makeatletter
\renewcommand{\theequation}{S\arabic{equation}}
\renewcommand{\thesection}{S\arabic{section}}
\renewcommand{\thefigure}{S\arabic{figure}}
\renewcommand{\bibnumfmt}[1]{[S#1]}
\renewcommand{\citenumfont}[1]{S#1}

\small
\section{Experimental details}
Yeast genetic manipulations were carried out as described in \cite{guthrie02}.  For the motor mutants, the cin8 and kip1 deletion strains were tested in several different ways: PCR for the presence of the KAN gene in the cin8 and kip1 locus; PCR for the absence of a wildtype copy of CIN8 and KIP1; and the most telling, a functional experiment: synthetic lethality of the cin8/kip1 double mutant, having one marked with NAT, the other with KAN. In crossing the cin8 strain to kip1, no spore was recovered that carried both mutations, of the approximately 60 tetrads tested, while all the other expected genotypes show up as expected, excluding the presence of a suppressor in either of the strains. Taken together, the PCR results verify that the deletions are present and that there is no wild type copy of either Kip1p or Cin8p present; and the crosses verify that these strains are synthetically lethal. 
Fig. S1 shows that when we remove either of the two types of kinesin-5 tetrameric motors, kip1 or cin8, using complete deletion of the gene, we find no statistically significant change in the observed spectrum. 

\begin{figure}[h]
\centering
  \includegraphics[width=0.95 \columnwidth]{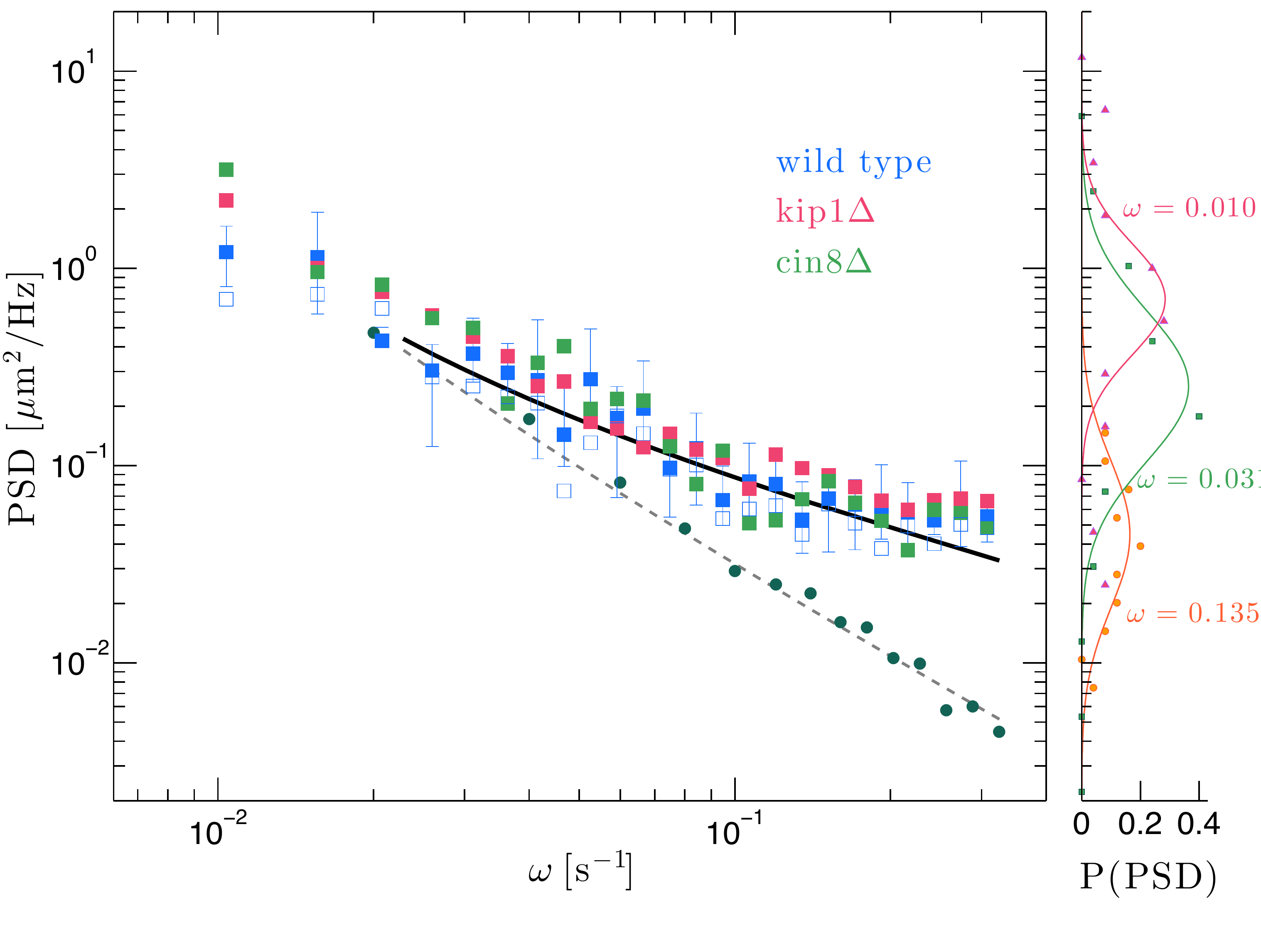}
  \caption{The {\it In vivo}  PSD determined in metaphase for separate populations of cells lacking the gene for one or the other of the two types of kinesin-5 tetrameric motors, Kip1 and Cin8, plotted together with the observed spectrum from unperturbed metaphase cells (Fig. 2) for comparison. }
  \label{fig:perturbations}
\end{figure}


\end{thebibliography}

\begin{thebibliography}{50}\frenchspacing

\bibitem{BranFCMWeitzJCB08} C. P. Brangwynne, G. H. Koenderink, F. C. MacKintosh, and D. A. Weitz, J. Cell Biol. {\bf 183}, 583 (2008).

\bibitem{Lau-PRL03} A.W. C. Lau, B. D. Hoffman, A. Davies, J. C. Crocker, and T. C. Lubensky, Phys. Rev. Lett. {\bf 91}, 198101 (2003).

\bibitem{Daisuke-Christoph-beadcells-2009} D. Mizuno, R. Bacabac, C. Tardin, D. Head, and C. F. Schmidt, Phys. Rev. Lett. {\bf 102}, 168102 (2009).

\bibitem{Fakhri2014} N.~Fakhri, A.~Wessel, C.~Willms, M.~Pasquali, D.~Klopfenstein, F.~MacKintosh,
  and C.~Schmidt, {Science}, {\bf 344}, 1031 (2014).
  
\bibitem{Daisuke-Science-2007} D. Mizuno, C. Tardin, C. F. Schmidt, and F. C. MacKintosh, Science, {\bf 315}: 370 (2007). 

\bibitem{Guo-2014} M.~Guo, A.~Ehrlicher, M.~Jensen, M.Renz, J.~Moore, R.~Goldman,
  J.~Lippincott-Schwartz, F.~Mackintosh and D.~Weitz, Cell, {\bf 158}: 822 (2014).
  
\bibitem{Gasser-PNAS04} K. Bystricky, P. Heun, L. Gehlen, J. Langowski, and S. M. Gasser, Proc. Nat. Acad. Sci. USA {\bf 101}, 16495 (2004).

\bibitem{LarsonBloom-2010} M. E. Larson, B. D. Harrison, K. Bloom, Biochimie {\bf 92}, 1741 (2010).

\bibitem{Weber-PNAS-2012} S. C. Weber, A. J. Spakowitz, and J. A. Theriot, Proc. Natl. Acad. Sci. USA {\bf 109}, 7338 (2012).

\bibitem{MacKintosh-PNAS-2012} F. C. MacKintosh, Proc. Natl. Acad. Sci. USA {\bf 109}, 7138 (2012).

\bibitem{DogteromMTforces} M. Dogterom and B, Yurke, Science {\bf 278}, 856 (1997).

\bibitem{MitchisonRef} P. Maddox, A. Straight, P. Coughlin, T. J. Mitchison, and E. D. Salmon, J. Cell Biol. {\bf 162}, 377 (2003).

\bibitem{StraightMurray-Science97} A. F. Straight, W. F. Marshall, J. W. Sedat, and A. W. Murray, Science {277}, 574 (1997).

\bibitem{Cross-irreversibleCellCycleStart} G. Charvin, C. Oikonomou, E. D. Siggia, and F. R. Cross, PLoS Biol. {8}, e1000284 (2010).

\bibitem{Boekeyeaststrain} C.~B. Brachmann, A.~Davies, G.~J. Cost, E.~Caputo, J.~Li, P.~Hieter and J.~D.
  Boeke, Yeast {14}, 115 (1998).

\bibitem{CSHL_manual} D. C. Amberg, D. J. Burke, and J. N. Strathern, {\em Methods in Yeast Genetics: A Cold Spring Harbor Laboratory Course Manual, 2005 Edition} (Cold Spring Harbor Laboratory Press, 2005).

\bibitem{KellerBusmotors} D.~Keller and C.~Bustamante, Biophys. J. {78}, 541 (2000).
  
\bibitem{tracking09} Y. Gao and M. Kilfoil, Optics Express {\bf 17}, 4685 (2009). 

\bibitem{ChristophPSD98} F. Gittes and C. F. Schmidt, Meth. Cell Biol. {\bf 55}, 129 (1998).

\bibitem{ActiveGelThyPRL} F. C. MacKintosh and A. J. Levine, Phys. Rev. Lett. {\bf 100}, 018104 (2008).

\bibitem{modelGardnerOdde} M. K. Gardner, C. G. Pearson, B. L. Sprague, T. R. Zarzar, K. Bloom, E. D. Salmon, and D. J. Odde, Mol. Biol. Cell {\bf 16}, 3764 (2005).

\bibitem{VeigelSchmidt-review_motors-2011} C. Veigel and C. F. Schmidt, Nat. Rev. Mol. Cell Bio. {\bf 17}, 244 (2009).

\bibitem{Khmelinskii-DevCell-2009} A. Khmelinskii, J. Roostalu, H. Roque, C. Antony, and E. Schiebel, Dev. Cell {\bf 17}, 244 (2009).

\bibitem{Levine-FCM-JPC09} A. J. Levine and F. C. MacKintosh, J. Phys. Chem. B {\bf 113}, 3820 (2009).

\bibitem{Landau} L.~D. Landau and E.~M. Lifshitz, {\em Statistical Physics Pt. 2} (Pergamon Press, Oxford, 1980), 2nd ed.

\bibitem{masonweitz} T. G. Mason and D. A. Weitz, Phys. Rev. Lett. {\bf 74}, 1250 (1995).

\bibitem{GittesPRL97} F. Gittes, B. Schnurr, P. D. Olmsted, F. C. MacKintosh, and C. F. Schmidt, Phys. Rev. Lett. {\bf 79}, 3286 (1997).

\bibitem{Schnurr1997} B.~Schnurr, F.~Gittes, F.~C. MacKintosh and C.~F. Schmidt, Macromolecules {\bf 30}, 7781 (1997).

\bibitem{microrheoDasguptaWeitz} B. R. Dasgupta, S. Tee, J. C. Crocker, B. J. Frisken, and D. A. Weitz, Phys. Rev. E {\bf 65}, 051505 (2002).
	
\bibitem{microrheo09} V. Pelletier, N. Gal, P. Fournier, and  M. L. Kilfoil, Phys. Rev. Lett. {\bf 102}, 188303 (2009).

\bibitem{Pernodet_lambdaDNA_xover97} N. Pernodet and B. Tinland, Biopolymers {\bf 42}, 471 (1997).

\bibitem{Larson_book} R. G. Larson  {\em The Structure and Rheology of Complex Fluids} (Oxford University Press, New York, 1999).

%
\bibitem{Discher-nucleusZimm07} J. D. Pajerowski, K. N. Dahl, F. L. Zhong, P. J. Sammak, and D. E. Discher, Proc. Natl. Acad. Sci. USA {\bf 104}, 15619 (2007).

\bibitem{SaundersHoyt_kinesinproteins92} W.~S. Saunders and M.~A. Hoyt, Cell {\bf 70}, 451 (1992).

\bibitem{SaundHoyt95} W.~S. Saunders, D.~Koshland, D.~Eshel, I.~R. Gibbons and M.~A. Hoyt, J. Cell Biol. {\bf 128}, 617 (1995).

\bibitem{Avunie-Masala2011} R.~Avunie-Masala, N.~Movshovich, Y.~Nissenkorn, A.~Gerson-Gurwitz, V.~Fridman, M.~K. {o}ivom\"{a}gi, M.~Loog, M.~A. Hoyt, A.~Zaritsky and L.~Gheber, J. Cell Sci. {\bf 124}, 873 (2011).

\bibitem{FridmanGheber2013} V.~Fridman, A.~Gerson-Gurwitz, O.~Shapira, N.~Movshovich, S.~Lak\"{a}mper, C.~Schmidt and L.~Gheber, J. Cell Sci. {\bf 126}, 4147 (2013).

\bibitem{OzalpOlsen_nanoATPlevelsYeast2010} V.~C. \"{O}zalp, T.~R. Pedersen, L.~J. Nielsen and L.~F. Olsen, J. Biol. Chem. {\bf 285}, 37579 (2010).

\bibitem{Ditzelmuller-ATPdepl83} G.~Ditzelm{\"u}ller, W.~W{\"o}hrer, C.~P. Kubicek and M.~R{\"o}hr, Arch. Microbiol. {\bf 135}, 63 (1983).

\begin{thebibliography}{1}
\bibitem{guthrie02} C.~Guthrie and G.~R. Fink, \emph{Methods in Enzymology}, Academic Press,
  (2002).
\end{thebibliography}
\end{document}